\documentclass[prl,twocolumn,floatfix,superscriptaddress]{revtex4-1}
\usepackage{dcolumn,amsmath}
\usepackage{graphicx}
\usepackage{bm}
\usepackage{hyperref}

\begin{document}

\title{Theoretical study of $^{173}$YbOH to search for the nuclear magnetic quadrupole moment}

\date{\today}

\begin{abstract}
A $CP$-violating interaction of the nuclear magnetic quadrupole moment (MQM) with electrons in the ytterbium monohydroxide molecule $^{173}$YbOH is considered. Both the MQM of the $^{173}$Yb nucleus and the molecular interaction constant $W_M$ are estimated. Electron correlation effects are taken into account within the relativistic Fock-space coupled-cluster method.
Results are interpreted in terms of the strength constants of $CP$-violating nuclear forces, neutron electric dipole moment (EDM), QCD vacuum angle $\theta$, quark EDMs and chromo-EDMs.
\end{abstract}

\author{D.E.\ Maison}\email{daniel.majson@mail.ru}
\affiliation{Saint Petersburg State University, 7/9 Universitetskaya Naberezhnaya, St. Petersburg, 199034, Russia}
\affiliation{Petersburg Nuclear Physics Institute  of National Research Centre
``Kurchatov Institute'', Gatchina, Leningrad District 188300, Russia}
\homepage{http://www.qchem.pnpi.spb.ru}
\author{L.V.\ Skripnikov}
\affiliation{Petersburg Nuclear Physics Institute named by B.P. Konstantinov of National Research Centre
``Kurchatov Institute'', Gatchina, Leningrad District 188300, Russia}
\affiliation{Saint Petersburg State University, 7/9 Universitetskaya Naberezhnaya, St. Petersburg, 199034, Russia}
\author{V. V. Flambaum}
\affiliation{School of Physics, The University of New South Wales, Sydney NSW 2052, Australia}
\affiliation{Johannes Gutenberg-Universit\"at Mainz, 55099 Mainz, Germany}

\maketitle

\section{Introduction}

The search for the effects of violation of the spatial parity ($P$) and time-reversal ($T$) symmetries of fundamental interactions is of importance to test modern extensions of the standard model~\cite{Safronova:18}. Due to the $CPT$ theorem, violation of the $T$ symmetry leads to violation of the $CP$, where $C$ is the charge conjugation. Understanding the nature of $CP$ violation is closely related to the bariogenesis problem \cite{sakharov1967violation}, which is important for cosmology and astrophysics.

The development of atomic and molecular spectroscopy methods already allows one to probe such effects at the energy scale of tens of TeV in experiments to search for $T,P$-odd effects 
%Victor
 produced by electron electric dipole moment (EDM), scalar-pseudoscalar nuclear-electron interaction ~\cite{Gorshkov:79}, 
 the dark matter candidates axion and relaxion \cite{stadnik2018improved},  etc. (see, e.g., Ref. ~\cite{ACME:18}).

It was shown in Ref. \cite{Flambaum:76} that the $T,P$-odd effects
%Victor
 produced by electron EDM rapidly grow in heavy atoms,  faster than $Z^3$, 
 where $Z$ is the charge of the nucleus. The most accurate atomic experiment to search for the electron EDM has been performed on a thallium atomic beam~\cite{Regan:02}. The next important stage was the experiment on the diatomic molecule YbF, where a slightly stronger limitation on the electron EDM was obtained: $|d_e|<1.1\times 10^{-27}\ e\cdot \textrm{cm}$ \cite{hudson2002measurement}. Further (about an order of magnitude) improvement was obtained on the ThO molecular beam experiment \cite{ThO}. Another type of experiment has been performed on  trapped molecular HfF$^+$ cations~\cite{cairncross2017precision}. In both cases the so-called $\Omega$-doublet structure of energy levels was employed. Such a level structure leads to the existence of closely spaced energy levels of opposite parity. The latter allows one to fully polarize molecules at very small external electric fields which simplifies the corresponding experiment. It also allows one to minimize some systematic effects~\cite{DeMille2001, Petrov:14, Vutha2009,Petrov:19a}. The best current limitation on the electron EDM, $|d_e|<1.1\cdot 10^{-29} e\times \textrm{cm}$, was obtained in the second generation of the ThO experiment \cite{ACME:18}. 
%ls
A very important feature of diatomic molecules with heavy atoms with respect to the heavy atoms is the existence of a very large effective electric field that interacts with the electron EDM. For example, in an external electric field of a few V/cm, the effective electric field in the working $^3\Delta_1$ electronic state of ThO achieves about 80 GV/cm \cite{Skripnikov:16b,Skripnikov:15a,Fleig:16}.
%end ls
 However, both improvement of the experimental technique and treatment of new systems are necessary to probe the new energy scale via the measurement of the T,P-odd effects~\cite{isaev2016polyatomic,kozyryev2017precision,Chubukov:19a}.

A recent suggestion made in Refs. \cite{isaev2016polyatomic,kozyryev2017precision} is to perform analogous experiments with linear triatomic molecules. With such systems it is expected to probe high-energy physics beyond the standard model in the PeV regime~\cite{kozyryev2017precision}.

Monohydroxides of alkaline-earth or earthlike-metal radicals are isoelectronic to the corresponding diatomic fluorides of the metals. These molecules, such as BaOH, RaOH, etc., are expected to have similar effective fields as their fluoride analogs BaF, RaF due to similar electronic structures. However, in contrast to the fluorides, they can have a much smaller energy gap between levels of opposite parity due to the $l$-doublet~\cite{kozyryev2017precision} structure in the low-lying excited vibrational states. This means that they can be fully polarized at small electric fields (which simplifies the experiment).
For example, YbF molecules were polarized by about 50\% at the  fields used~\cite{Hudson:11a}. This feature of simple polarization of such triatomic molecules is similar to that in the ThO molecule. However, one can find triatomic molecules which can also be laser cooled, which is not possible for the latter case. The YbOH molecule is such a triatomic molecule proposed for the electron EDM experiment~\cite{kozyryev2017precision}. In the experiment the ground electronic state $^2\Sigma$ will be used. The electronic configuration of this state corresponds to one unpaired electron over closed shells, which is occupying the hybrid $sp$ shell of the Yb atom, similar to the YbF case \cite{Titov:96b,quiney1998hyperfine}.

Recently, there were several theoretical studies of the YbOH molecule as a candidate for search of the electron electric dipole moment \cite{gaul2018ab,denis2019enhancement,prasannaa2019enhanced}. In the present paper we consider this molecule to search for another type of the $T,P$-violation source, the nuclear magnetic quadrupole moment (MQM). 
%Victor
Atomic EDM and $T,P$-violation effects in molecules produced by the nuclear MQM increase with a nuclear charge faster than $Z^2$ \cite{Sushkov:84}, therefore, at least one heavy atom, such as Yb,  is needed. An additional enhancement appears in deformed nuclei where MQM has a collective nature \cite{F94}.
MQM can be nonzero only for nuclei with spin $I\ge 1/2$. Ytterbium has one such stable isotope $^{173}$Yb, 
%Victor
a deformed nucleus, and its enhanced collective  MQM leads to the $T,P$-odd energy shift in the $^{173}$YbOH molecule -- MQM interacts with the gradient of the magnetic field produced by electrons.  We study this interaction and calculate the molecular constant which connects the possible experimental energy shift with the $^{173}$Yb nucleus MQM that will be required for the interpretation of the experiment. Measurement of the nonzero nuclear MQM value would indicate the presence of 
%Victor
nuclear $T,P$-odd forces and nucleon electric dipole moments, and this would have great consequences for the unification theories  predicting a $CP$ violation.

%ls
%\section{Theory}
\section{Electronic structure parameter}
%end ls
The $T,P$-odd interaction of the nuclear magnetic quadrupole moment with electrons is given by the following Hamiltonian~\cite{GFreview},
\begin{equation} \label{HmqmOperator}
    H_{MQM} 
    =
    -\frac{M}{2I\left(2I-1\right)} T_{i,k} \cdot \frac{3}{2} \frac{\left[\boldsymbol{\alpha}\times \mathbf{r}\right]_i r_k}{r^5},
\end{equation}
where $T_{i,k}=I_i I_k+ I_k I_i -\frac{2}{3}I(I+1)\delta_{ik}$, $I$ is the nuclear spin of $^{173}$Yb, $M$ is the magnetic quadrupole moment of the $^{173}$Yb nuclei, $\boldsymbol{\alpha}$ are Dirac matrices and $\mathbf{r}$ is the electron radius-vector with respect to the heavy atom nucleus under consideration. 
 
The electronic part of the Hamiltonian (\ref{HmqmOperator}) is characterized by the molecular constant $W_M$~\cite{Sushkov:84,Kozlov:95},
which is given by
\begin{equation}
\label{WmME}
    W_M
    =
    \frac{3}{2 \Omega}
    \langle \Psi | \sum\limits_i 
    \left(
        \frac{\boldsymbol{\alpha}_i\times\mathbf{r}_i}
             {r_i^5}
    \right)_\zeta r_\zeta | \Psi \rangle,
\end{equation}
where $\Omega$ is the projection of the total electronic angular momentum $\mathbf{J}^e$ on the molecular axis, $\Psi$ is the electronic molecular wavefunction, andthe sum index $i$ is over all the electrons. A $W_M$ constant is required for interpretation of the experimental data in terms of the nuclear MQM. The  ground electronic state of the YbOH molecule has $\Omega = 1/2$. 

\section{Nuclear magnetic quadrupole moment for Yb}

The nucleus $^{173}$Yb is deformed, and we base our calculations on the results of the MQM calculations in the Nilsson model presented in Ref. \cite{lackenby2018time}. Summation over nucleons  gave the following result for the  $^{173}$Yb collective MQM \cite{lackenby2018time}:
\begin{equation}
\label{M}
M= 14 M^p_0 + 26 M^n_0,
\end{equation}
where  $M^p_0$  and  $M^n_0$ are the single-particle matrix elements for protons and neutrons  which depend on the form of the $T,P$-odd interaction. We start from a contact $T,P$-odd nuclear  potential 
\begin{equation}
\label{VTP}
V^{TP}_{p,n}= \eta_{p,n}  \frac{G} {2^{3/2} m_p} (\sigma \cdot \nabla \rho),
\end{equation}
acting on the valence nucleon. Here $\eta_{p,n}$ is the dimensionless strength constant,  $\rho$ is the total nucleon number density, $G$ is the Fermi constant, and $m_p$ is the proton mass. Using Eq. (\ref{M}) and values of $M^p_0$  and  $M^n_0$  from  Refs. \cite{FDK14,lackenby2018time} we obtain
\begin{equation}
\label{Meta}
M= (2 \eta_n - \eta_p) \times 10^{-33} e \cdot  \text{cm}^2 + (0.6 d_n + 0.3 d_p) \cdot 10^{-12}  \text{cm},
%\frac{\hbar}{m_p c} ,
\end{equation}
where $d_n$ and $d_p$ are neutron and proton electric dipole moments.
%$\lambda_p=\hbar /m_pc=2.10 \cdot 10^{-14}$ cm.
The $T,P$- odd nuclear potential Eq. (\ref{VTP}) is dominated  by the neutral $\pi_0$ exchange between the nucleons and the strength constants $\eta$ may be expressed in terms  of $\pi NN$ couplings (see details in  Ref. \cite{FDK14}):
\begin{align}
\eta_{n} = -\eta_{p} \approx 5\times 10^{6}g\left(\bar{g}_1 + 0.4 \bar{g}_2\ - 0.2\bar{g}_0\right) ,
\end{align}
where $g$ is the strong $\pi NN$ coupling constant  and  $\bar{g}_0\,, \bar{g}_1\,, \bar{g}_2$ are  three $T$-,$P$-odd $\pi NN$ coupling constants, corresponding to the different isotopic  channels.  Substitution of these $\eta_{n,p}$ into Eq. (\ref{Meta}) gives:
\begin{eqnarray}
\label{Mg}
\nonumber
M= g \left(1.5 \bar{g}_1 + 0.6 \bar{g}_2\ - 0.3\bar{g}_0\right) \times 10^{-26} e \cdot  \text{cm}^2 \\
+ (0.6 d_n + 0.3 d_p) \cdot 10^{-12}  \text{cm},
%\frac{\hbar}{m_p c} ,
\end{eqnarray}
Constants of the $T,P$-odd $\pi NN$ interaction  $\bar{g}$ and nucleon EDMs may be expressed in terms of more fundamental $T,P$- violating parameter,  QCD constant  $\bar{\theta}$, or EDM  $d$ and chromo-EDM $\tilde{d}$ of $u$ and $d$ quarks \cite{theta,PospelovRitzreview}:

$$g\bar{g}_0(\bar{\theta}) =  -0.37 \bar{\theta}$$
$$d_n=-d_p=1.2 \cdot 10^{-16} \bar{\theta} \cdot  e \cdot \text{cm}$$
$$g\bar{g}_0(\tilde{d}_u, \tilde{d}_d) = 0.8\times 10^{15} \left(\tilde{d}_u +\tilde{d}_{d}\right) \ \text{cm}^{-1} $$
$$g\bar{g}_1(\tilde{d}_u, \tilde{d}_d) = 4\times 10^{15} \left(\tilde{d}_u  - \tilde{d}_{d}\right) \ \text{cm}^{-1}$$
$$d_{p}(d_u, d_d, \tilde{d}_u, \tilde{d}_d) = 1.1e\left(\tilde{d}_u + 0.5\tilde{d}_{d}\right) + 0.8 d_u - 0.2d_d $$
$$d_{n}(d_u, d_d, \tilde{d}_u, \tilde{d}_d) = 1.1e\left(\tilde{d}_d + 0.5\tilde{d}_{u}\right) - 0.8 d_d + 0.2d_u$$

The substitutions to Eq.  (\ref{Mg}) give the following results for MQM:
\begin{align}  \label{Md}
M(\bar{\theta}) \approx 1. \cdot 10^{-27} \, \bar{\theta} e \cdot  \text{cm}^2 \,, \\
%\begin{split}
\label{MdD}
M(\tilde{d}) \approx   0.6  \times 10^{-10 }  \left(\tilde{d}_u  - \tilde{d}_{d}\right) e \cdot  \text{cm}   
%\end{split} 
\end{align}

Using updated results \cite{Yamanaka2017,Vries2015}:
    \begin{align}
       \bar{g}\bar{g}_0 &= -0.2108 \bar{\theta} \, , \\
       \bar{g}\bar{g}_1 &= 46.24 \cdot 10^{-3} \bar{\theta} \, ,
    \end{align}
we obtain a slightly larger value of $M(\bar{\theta})$ which is still approximately given b any estimate in Eq.(8).

\section{Electronic structure calculation details}

The main calculations were performed within two Gaussian-type basis sets:
the LBas basis set consists of the uncontracted Dyall's all-electron quadruple-zeta (AE4Z) \cite{Dyall:2016} for the Yb atom and the augmented correlation-consistent polarized triple-zeta (aug-cc-PVTZ-DK) basis set \cite{Dunning:89,Kendall:92} for the oxygen and hydrogen atoms;
the SBas basis set consists of the uncontracted Dyall's all-electron double-zeta, AE2Z, basis set \cite{Dyall:2016} for the Yb atom and the augmented correlation-consistent polarized double-zeta (aug-cc-PVDZ-DK) basis set \cite{Dunning:89,Kendall:92,de2001parallel} for oxygen and hydrogen atoms. Also, the uncontracted AE3Z~\cite{Dyall:2016} basis set for Yb was used for the analysis of basis set convergence.

For the main contribution to the $W_M$ parameter the LBas basis set was used. In this calculation $1s\dots 2p$ electrons of Yb were excluded from the correlation treatment. Calculation was performed within the relativistic Fock-space coupled-cluster method with single and double amplitudes (FS-CCSD) using sector (0,1) of the Fock space. In this calculation sector (0,0) corresponded to the YbOH$^+$ cation. The energy cutoff for virtual orbitals was set to 450 hartree in the correlation treatment. The correlation contribution of $1s\dots 2p$ electrons to $W_M$ was obtained within the SBas basis set as a difference between the all-electron result and 69-electron one (with frozen $1s\dots 2p$ electrons). In the all-electron calculation the cutoff energy for virtual orbitals was set to 10500 hartree. In Ref.~\cite{Skripnikov:17a} it was demonstrated that such an energy cutoff is important to ensure including functions that describe spin-polarization effects for inner-core electrons. Also the importance of the high energy cutoff was extensively analyzed in Ref.~\cite{Skripnikov:15a} for the correlation contribution of the outer-core electrons.

Correlation calculations were performed using the {\sc dirac15} code \cite{DIRAC15}. To compute the matrix elements (\ref{WmME}), the code developed in Ref.~\cite{Skripnikov:17b} was used.

\section{Results and discussion}
Table \ref{TResult1} gives values of $W_M$ calculated within different basis sets. One can see a good convergence with respect to the basis set size.
In particular, values obtained with the AE3Z and AE4Z basis sets of Yb differ only by about 0.6\%. 

\begin{table}[!h]
\caption{
Dependence of the calculated value of the $W_M$ parameter for YbOH on different basis sets within the relativistic FS-CCSD approach; $1s^22s^22p^6$ electrons of Yb were frozen.
}
\label{TResult1}
\begin{tabular}{lll}
\hline
\hline
Basis for Yb $ \ \ \ \ $  & basis for  O and H  $ \ \ \ \ $  & $W_M$, 10$^{33} \frac{\textrm{Hz}}{e\cdot \textrm{cm}^2}$ \\
\hline
AE2Z & aug-cc-pVDZ-DK & -1.040\\
AE3Z & aug-cc-pVDZ-DK & -1.063\\
AE3Z & aug-cc-pVTZ-DK & -1.060\\
AE4Z & aug-cc-pVTZ-DK & -1.066\\
\hline
\hline
\end{tabular}
\end{table}

Table \ref{TResult2} presents the final value of $W_M$ as well as its contributions. The final value of $W_M$ is $-$1.07(5)$\times 10^{33}$Hz/($e$~$\cdot$~cm$^2$). The contribution of the Gaunt interaction is obtained as a difference between the values of $W_M$ calculated at the Dirac-Fock-Gaunt and Dirac-Fock levels, which is about -0.3\% (see Table \ref{TResult2}). 
One should note a considerable contribution of the correlation effects: It is more than 30\% of the total value.
%dm
The estimated theoretical uncertainty of the $W_M$ value should be included in the value of MQM when the experiment will be performed.
%end dm

\begin{table}[!h]
\caption{
Calculated value of the $W_M$ parameter for YbOH.
}
\label{TResult2}
\begin{tabular}{ll}
\hline
\hline
%Victor  {cm}^2 instead of {cm}
Method $\ \ \ \ \ $& $W_M$, 10$^{33} \frac{\textrm{Hz}}{e\cdot \textrm{cm}^2}$ \\
\hline
DHF & -0.736 \\
FS-CCSD & -1.066 \\
%\hline
%\multicolumn{2}{c}{Corrections} \\
%\hline
+Inner electrons ($1s..2p$) & -0.011 \\
+Gaunt & +0.003\\
%\hline
\\
Total & -1.074 \\
\hline
\hline
\end{tabular}
\end{table}

The obtained value is close to the corresponding value for the YbF molecule: $W_M=$-1.3$\times 10^{33}$Hz/($e$~$\cdot$~cm$^2$)~\cite{Titov:96b,quiney1998hyperfine}. One can also compare $W_M$(YbOH) with $W_M$ for other molecules on which experiments to search for $T,P$-odd effects are conducted at present. In particular, $W_M$(YbOH) is more than twice larger than the same characteristic in the HfF$^+$ molecular cation, $W_M=$0.494$\times 10^{33}$Hz/($e$ $\cdot$ cm$^2$) \cite{Skripnikov:17b,Fleig:17}; it is close to that in the ThO molecule, $1.10 \times 10^{33}$Hz/($e$~$\cdot$~cm$^2$)~\cite{Skripnikov:14a,Skripnikov:14aa} and about twice larger than in the ThF$^+$ cation, $0.59~\times~ 10^{33}$Hz/($e$~$\cdot$~cm$^2$) \cite{Skripnikov:15b} 
\footnote{Note that the $W_M$ value given in Ref.~\cite{Skripnikov:15b} should be corrected by the factor of $2/3$ as explained in \cite{Skripnikov:14aa}}.

The final energy shift produced by the interaction of the nuclear MQM with electrons described by the Hamiltonian given by Eq. (\ref{HmqmOperator}) can be represented in the following form~\cite{Skripnikov:14c},
\begin{equation}
\label{EnergyShift}
     \delta(J,F,\Omega)=C(J,F,\Omega) W_M M,
\end{equation}
where $F$ is the total angular moment and $J$ is the total moment less nuclear spins. $C(J,F,\Omega)$ depends on the actual experimental conditions: the hyperfine sublevel and external electric field used. 
By an order of magnitude this factor can be estimated as 0.1 \cite{Skripnikov:14a,Petrov:18b}.
Thus, taking into account this value, Eqs.(\ref{EnergyShift}), (\ref{Md}), and (\ref{MdD}), as well as the calculated value of $W_M$, one can express the energy shift in terms of the fundamental $CP$-violating physical quantities $\bar{\theta}$ and $\tilde{d}_u, \tilde{d}_{d}$:
\begin{equation}
|\delta(\theta)| \approx 10\times10^{10}\bar{\theta}\cdot\mu\text{Hz},
\end{equation}
\begin{equation}
|\delta(\tilde{d}_u -\tilde{d}_{d})| \approx 6\times\frac{10^{27}(\tilde{d}_u -\tilde{d}_{d})}{\text{cm}}\cdot\mu\text{Hz}.
\end{equation}
The current limits on $|{\tilde \theta}|$ and $|{\tilde d_u}{-}{\tilde d_d}|$ ($|{\tilde \theta}| < 2.4 \times 10^{-10}$, $|{\tilde d_u}{-}{\tilde d_d}|<6 \times 10^{-27} $~cm, see Ref.~\cite{Hg}) correspond to the shifts $|\delta| < 24~\mu$Hz and $36~\mu$Hz, respectively. 
These values are already of the same order of magnitude as the current accuracy achieved in measurements of the energy shift produced by the electron EDM in $^{232}$ThO~\cite{ACME:18}.
In Ref.~\cite{kozyryev2017precision} it was suggested that by using the YbOH molecule the sensitivity to the electron EDM can be increased by four orders of magnitude above the that obtained in Ref. ~\cite{ThO} (and, consequently, three orders with respect to Ref. ~\cite{ACME:18}). Thus one can expect that similar experiment on $^{173}$YbOH can significantly improve the limits on the $|{\tilde \theta}|$ term and on the difference of the quark chromo-EDMs $|{\tilde d_u}{-}{\tilde d_d}|$.
%end ls

%\section{Conclusion}

\section{Acknowledgments}
Electronic structure calculations were supported by the Russian Science Foundation Grant No. 18-12-00227. Calculations of the $W_M$ matrix elements were supported by the foundation for the advancement of theoretical physics and mathematics ``BASIS'' grant according to the Research Project No. 18-1-3-55-1.
%Victor
%ls: if possible (our grants requires direct separation of acknowledges)
% V.F. acknowledges support from
Nuclear structure calculations were supported by
%end ls
 the Australian Research Council and Gutenberg Fellowship. 
Electronic structure calculations in the paper were carried out using resources of the collective usage centre "Modeling and predicting properties of materials» at NRC "Kurchatov Institute" - PNPI.
%Victor
The authors would like to express special thanks to the Mainz Institute for Theoretical Physics (MITP) for its hospitality and support
and to Dr. A.N. Petrov for valuable discussions.

\bibliographystyle{apsrev}

\end{document}